# Magnetic characterization of undoped and 15%F doped LaFeAsO and SmFeAsO compounds

M.R. Cimberle[1*], F.Canepa[1,3], M.Ferretti[2,3], A. Martinelli[2], A.Palenzona[2,3], A.S. Siri[2,4], C. Tarantini[5], M. Tropeano[2,4], C. Ferdeghini[2]

[1]*CNR-IMEM via Dodecaneso 33, 16146 Genova – Italy*

[2]*CNR-INFM-LAMIA Corso Perrone 24, 16152 Genova – Italy*

[3]*Dipartimento di Chimica e Chimica Industriale, Università di Genova via Dodecaneso 31, 16146 Genova – Italy*

[4]*Dipartimento di Fisica, Università di Genova via Dodecaneso 33, 16146 Genova – Italy*

[5]*National High Magnetic Field Laboratory, Florida State University, Tallahassee, Florida 32310, USA*

## Abstract

In this paper the magnetic behaviour of undoped and 15% F doped SmFeAsO (Sm-1111) and LaFeAsO (La-1111) samples is presented and discussed. Magnetization measurements are not a simple tool to use for the characterisation of the new family of Fe-based superconductors because magnetic impurities can be easily formed during the preparation procedure and may affect the magnetic signal. In spite of this problem bulk magnetization measurements, properly treated, may give very useful information. In the undoped samples we gathered the main aspects of the physical behavior of the 1111 phase, i.e. the onset of the Spin Density Wave (SDW), the antiferromagnetic ordering at the Sm sublattice and the susceptibility increase with increasing temperature above the SDW temperature, and, in addition, we were able to estimate the Pauli contribution to susceptibility and therein the Wilson ratio both for LaFeAsO and SmFeAsO compounds, and the amplitude of the jump at the SDW temperature. In the doped samples, while the presence of magnetic signals due to impurities is dominating in the normal state, the superconducting behavior may be clearly observed and studied. In particular, in the Sm-1111 superconducting sample the coexistence-competition between superconductivity and antiferromagnetic ordering of the Sm ions was clearly observed.

*corresponding author:

Maria Roberta Cimberle

IMEM-CNR c/o Dipartimento di Fisica dell'Università di Genova-via Dodecaneso 33  16146 Genova Italy

tel:+390103536448 fax:+390103622790

e-mail: cimberle@fisica.unige.it

## 1. Introduction

The recent discovery of superconductivity at temperatures up to $T_c$=26K in the iron oxipnictide LaFeAs($O_{1-x}F_x$) [1] has stimulated a lot of work in this field. In a short period of time different approaches were attempted to increase the transition temperature. On the one hand, La was substituted with other Rare Earth (RE) with smaller ionic radius (such as Pr, Nd, Ce, Sm, Gd) in order to induce chemical pressure [2,3,4]; on the other hand, the optimal doping for the superconducting phase was studied and obtained with a variety of different techniques, i.e. F substitution at the O site[5], O deficiency [6], and partial substitution on the RE site with bi-and tetravalent cationic species [7,8]. By combining these approaches, $T_c$ was increased up to the considerable value of about 55K [2,3,4,7,9]. Simultaneously, many different structural and physical characterizations were performed, together with a remarkable theoretical effort devoted to the comprehension of the superconductivity mechanism and of the possible coexistence of superconductivity and magnetism.

These materials present a layered structure with Fe-As layers alternating to RE-O ones. The parent compound displays a structural distortion from tetragonal to orthorhombic crystal symmetry coupled with a Spin Density Wave (SDW) antiferromagnetic order which occurs in the Fe sublattice; this ordering is reflected in all physical properties (i.e. resistivity, specific heat, Hall effect, and far infrared reflectance), which all suggest the opening of an energy gap [9].

After doping, the antiferromagnetic ordering is suppressed as well as the structural distortion, and the compound exhibits a metallic behaviour down to the superconducting transition temperature. Whether magnetic fluctuations are involved in the development of superconductivity is an open issue that could give new insight on the analogous problem related to high $T_c$ superconductors. Moreover, if and what type of magnetism is present in the doped samples, both in the normal and superconducting state, is the object of a lively debate. In addition, when La is substituted with other RE (Pr, Nd, Sm, and Gd) another source of magnetism is introduced, although in the RE-O (charge reservoir) layers, and not in the superconducting (Fe-As) ones.

The study of the oxipnictides by magnetic measurements, commonly used to characterize new materials, is made complicated by the simultaneous presence of many magnetic signals due to: 1) magnetic impurities that, even if in small amounts, can give a more or less large signal that is added to the magnetic behavior of the studied phase, 2) the diamagnetism of the superconducting phase and 3) the presence of a magnetic Rare Earth that substitutes La, whose magnetic signal may be dominant.

For the last reason the feature that, in the undoped samples, marks the antiferromagnetic ordering due to the onset of the SDW is rarely shown in the literature by magnetisation data. Recently this feature has been extracted from magnetic measurements on Ce, Pr and Nd based oxipnictides by subtraction of the paramagnetic signal due to the RE sublattice from the total signal [10] .

In this paper we present the structural and magnetic characterization of both undoped and 15%F doped SmFeAsO and LaFeAsO samples. In the undoped samples, after the subtraction of a ferromagnetic background signal due to the impurities, all the important characteristics of the 1111 phases both for la and Sm based compounds were observed: in particular the paramagnetic behaviour of the Sm ions sublattice was fitted and this allowed to value both the amplitude of the step at the Spin density wave onset, both the value of the Pauli susceptibility for the phase. In the doped samples, on the other side, paramagnetic or antiferromagnetic impurities are present, which make uncertain to extract the magnetic behavior of the samples in the normal state. However important information regarding the interplay between superconductivity and antiferromagnetism of Sm ions sublattice is obtained and discussed.

## 2. Samples preparation and characterization

Both undoped and 15% F doped SmFeAsO and LaFeAsO were prepared in two steps: 1) synthesis of REAs starting from pure elements in an evacuated pyrex tube at a maximum temperature of 550°C for five days; then 2) synthesis of the oxypnictide in tantalum crucible in evacuated quartz tube by reaction of REAs with stoichiometric amounts of Fe, $Fe_2O_3$ and, for the doped compounds, $FeF_2$. The sintering step was 30 h at 1250°C for Sm phases and at 1150°C for La phases. Details concerning the preparation are reported in [11].

The samples were characterized by X-ray powder diffraction (XRPD) and by Scanning Electron Microscopy (SEM) equipped with an energy dispersive X-ray microprobe (EDS). Fig.1 shows the XRPD patterns of all the samples. The XRPD analysis confirms the formation of both REFeAsO and $REFeAs(O_{0.85}F_{0.15})$ phases with RE=Sm and La. Some spurious peaks are present in the doped samples, originated by the corresponding rare earth oxy-fluorides. In Sm undoped sample no impurity phase was detected by X-ray diffraction analysis, whereas SEM-EDS analysis indicates the presence of FeAs in traces. In the doped sample by Rietveld refinement of X-ray diffraction data the presence of both SmOF (4.6%) and FeAs(5.6%) was evaluated. Both undoped and doped Sm samples were constituted of connected micrometric crystals, as SEM observation revealed. After F-substitution the cell size contracts, with a more enhanced decrease along the c-axis.

In LaFeAsO sample metallic iron is detected by SEM-EDS analysis, inside which some As is dissolved (5%). On the other side the presence of both LaOF and $Fe_2As$ as secondary phases is detected in the doped sample. Both pure and doped samples are mainly constituted by aggregated microcrystals of the oxy-pnictide, similarly to what observed in Sm-1111 samples. In conclusion the doped samples show the presence of the rare earth oxy-fluoride, together with iron arsenide compounds. In any case more than 90% of the samples here presented corresponds to the right oxipnictide phase.

The resistivities of the La and Sm based samples, normalized to their values at 300K, are reported in Fig. 2 a) and b) respectively. The parent samples exhibit well defined maxima at the formation of the SDW. These maxima are at T= 147K and T=159 K (while the $d\rho/dT$ maxima are at T= 131 K and T=138 K) for Sm and La-1111 phases respectively. Below the SDW temperature the resistivity decreases. In Sm-1111 phase the resistivity decrease continues down to the lowest temperature we reached (400 mK): below about 6 K an enhanced drop is clearly visible. In the following we will correlate this drop to the susceptibility measurements and, in particular, to the establishment of the antiferromagnetic ordering of the Sm ions sublattice. In La-1111 phase below about 30 K a slight resistivity increase is present: such behaviour has been attributed to carrier localization arising from SDW gap [12]. The further decrease that we observe below 10 K is difficult to explain: we cannot exclude that can be due to non percolative superconductivity in a small sample region, where some oxygen deficiency is present .

The doped samples are characterized by a nearly linear decrease in resistivity with temperature. For LaFeAs($O_{0.85}F_{0.15}$) sample it results $T_c$ =26.5 K if defined as the onset of resistivity and $T_c$ =21K if estimated from the maximum of $d\rho/dT$: here the superconducting transition width is about 10 degrees. For SmFeAs($O_{0.85}F_{0.15}$) sample with the same criteria we obtained $T_c$ =53.5 K and  $T_c$ =51.5 K respectively.

All the magnetic measurements were performed with a SQUID magnetometer (MPMS by Quantum Design).

### 3. Magnetic measurements

### 3.1 Undoped samples

The molar susceptibility of both LaFeAsO and SmFeAsO samples, measured from 2K up to 360K, is shown in Fig.3. In these measurements, after a Zero Field Cooling (ZFC) procedure, a field of 30kOe was applied.

The main features for LaFeAsO sample (lower curve) can be summarized as follows: i) at T=152K a clear decrease in the susceptibility is observed; ii) at lower temperatures a minimum is observable followed by an increase in susceptibility down to the lowest temperature we reached; iii) at temperatures greater than 152 K, susceptibility increases with increasing temperature and a maximum at T ~ 350K is seen. This behavior has been already reported in the literature [5,13]. The susceptibility drop at T=152K is due to the SDW onset and the related antiferromagnetic ordering of Fe ions, while the susceptibility increase above 152 K is attributed to non conventional antiferromagnetic correlations ( pseudo-gap behaviour) of Fe ions in the Fe-As layers [14,15]. The susceptibility increase at low temperature is probably due to the presence of traces of spurious phases. The same hypothesis must be made for the faint peak at T=350K, a temperature rarely covered in measurements reported in the literature. In our opinion, it marks the presence of the phase $Fe_2As$, which it is known to have an antiferromagnetic transition right at T~350K [16]. We remark that neither X-ray diffraction nor SEM analyses gave an indication of the presence of $Fe_2As$ in this undoped sample: the magnetic measurement is a more sensitive instrument in this case. We suggest that an extension of the temperature range above room temperature could give a useful indication of the presence of $Fe_2As$.

For the SmFeAsO sample (upper curve) the main features are the following: i) at T~134 K a local maximum, related to the SDW onset is present; ii) at T=6 K there is a very sharp cusp that we attribute to the antiferromagnetic ordering of the Sm sublattice, in agreement with the sharp peak in the specific heat observed by Tropeano et al.[17] at 4.6 K in the same sample, and by Ding et al at 4.6K [18] in a similar sample, and iii) a Curie-Weiss like behaviour, related to the presence of paramagnetic Sm, is present. It is clear that the presence of the magnetic signal from Sm is dominating, and hides the possible presence of an increase in magnetization with temperature similar to that observed in La samples above the SDW transition.

Finally we point out that, for both the La and Sm-1111 samples, a small bump is present at about 40K. Such anomaly has been already observed in literature [13], and usually attributed to the presence of some magnetic impurity. The fact that it happens at the same temperature both in La and Sm-1111 samples makes difficult to attribute it to a onset of superconductivity for oxygen deficiency in some part of the sample.

We remark that for both LaFeAsO and SmFeAsO samples the temperature dependent susceptibility signal is superimposed on a temperature-independent background. We argue that part of this magnetic background is due to the ferromagnetic impurities present in our samples. This can be seen from magnetization versus field measurements at T=5K and T=300K up to 50 kOe, displayed in the insets of Fig.3 for both the samples. Here a ferromagnetic signal, sharply increasing at low field and saturating around 10 kOe, is present in the two compounds. In SmFeAsO, in addition, there is a clear linear increase in magnetisation above 10 kOe, due mainly to the paramagnetic contribution of the Sm ions (and also to the Pauli paramagnetic signal), while for LaFeAsO a very small field dependence is observed, and the curves at T=5K and T=300K nearly overlap. Since the saturation values of the ferromagnetic components for each sample are nearly the same at T=5K and 300K, we know that the ferromagnetic impurities have a transition temperature higher than the room temperature. The extrapolated magnetization values are 62 and 55 emu/mol at T= 5K and 300K for La, and 80 and 70 emu/mol at T= 5K and 300K for Sm. On the basis of these extrapolated magnetisation data, a linear behavior of the thermal dependence of the ferromagnetic impurity magnetisation has been hypothesized ($M(T) = 62.1 - 0.024 \cdot T$ and $M(T) = 80.2 - 0.034 \cdot T$ for LaFeAsO and SmFeAsO samples, respectively,) and subtracted from the measured curves, yielding the data of molar susceptibility shown in Fig.4.

These measurements represent well the magnetic behaviour of the 1111 phases. Moreover, following the idea proposed in [10], we fitted the susceptibility data (between 7 and 100K) of Sm-1111 phase with a Curie-Weiss type behavior plus a temperature independent term. Unlike in [10] we chose to fit the low temperature data, under the reasonable hypothesis that are dominated, for the temperature dependent part, by the magnetic signal of the Sm sublattice.

Therefore we wrote $\chi = \chi_0 + \chi_{Sm}$  (1)

The $\chi_0$ term covers all the different temperature independent contributions (Pauli paramagnetism and Landau diamagnetism of the conduction electrons, high frequency contribution, diamagnetic contributions arising from nuclei and electronic inner shells…).

It is well known that the Sm presents an anomalous not Curie-Weiss susceptibility behavior in most of its compounds. The successful explanation of this fact [19] is that in the $Sm^{3+}$ ion the J multiplet intervals are comparable with $kT$: so, not only the J = 5/2 ground state is occupied but also the first excited level J=7/2. As a consequence, the sum of the J levels, carried out up to the second term and taking the Boltzmann temperature factor into account, gives the relationship:

$$\chi_{Sm} = x \cdot \frac{C1}{(T - \vartheta_{Sm})} + (1 - x) \frac{C2}{(T - \vartheta_{Sm})} \cdot e^{(-\frac{\Delta}{T})}$$  (2)

where $x$ represents the fractional occupation of the two states, $\theta_{Sm}$ is the paramagnetic Curie temperature, C1 =0.0903 emu K/mol and C2=1.3452 emu K/mol are the calculated Curie constant related to the two J states, and $\Delta$ =E/k is the difference in temperature between the two states.

As parameters of the fit we obtained $\chi_0$ = 0.97×10⁻³ emu/mol , $x$ =0.84, $\theta_{Sm}$ = -59 K and $\Delta$ = 600 K. The result is presented in Fig.4 as a dotted line. We note that the energy gap is in good agreement with the value 650 K (= 56 meV) obtained from heat capacity measurements on SmFeAsO [20]. By subtracting from the experimental data at high temperature ( T>100 K) the $\chi$ values of the best fit, we obtained the intrinsic susceptibility contribution of the Fe-As layers, which is shown in the inset of Fig. 4 (upper right side). A clear jump at $T_{SDW}$ is emphasized. Its value is about 7×10⁻⁵ emu/mol. The LaFeAsO susceptibility jump has about the same value (6×10⁻⁵ emu/mol) and McGuire and co.[10] obtained comparable values for the other REFeAsO, *i.e.* ~5×10$^{-5}$ emu/mol for CeFeAsO and ~4×10$^{-5}$ emu/mol for PrFeAsO and NdFeAsO. This suggests a similar value of the susceptibility jump at the SDW ordering for all the RE, and supports the correctness of the estimate of the Sm sublattice susceptibility we did.

Let us discuss the relatively high positive value of $\chi_{0=}$ 0.97$^{.}$10$^{-3}$ emu/mol we obtained from the fit. $\chi_0$ is mainly ascribed to the Pauli positive contribution arising from non interacting band electrons $\chi_P = 2N(E_F)\mu_B^2$ where N(E$_F$), the density of states at the Fermi level, is proportional to the effective mass *m\** of the electrons through

$$N(E_F) = \frac{(2m^*)^{\frac{3}{2}} E_F^{\frac{1}{2}}}{4\pi^2} \qquad (3)$$

For normal metals $\chi_0$ is around 10$^{-5}$÷10$^{-6}$ emu/mol, which corresponds to an effective mass m* comparable with the mass of the free electron. Here $\chi_0$ is about one hundred times higher. So, the possibility that 4f electrons of Sm can hybridize with the conduction electrons, giving a strong enhancement of carrier effective mass (*heavy fermion state*), can be taken into account. Values of $\chi_0$ as high as this one or more were found in typical heavy fermion or concentrated Kondo systems: for example for CeCu$_2$Si$_2$ $\chi_0$=6.5 $^{.}$10$^{-3}$ emu/mol [21] and for CeAl$_3$, $\chi_0$=3.6$^{.}$10$^{-3}$ emu/mol [22]. Recent measurements of heat capacity on SmOFeAs gave a high value of the electronic coefficient $\gamma$ of the specific heat: on the same sample here presented it turned out to be $\gamma$ = 42 mJ/K$^2$ mol [17]. The knowledge of $\gamma$ and $\chi_0$ allows one to calculate the Wilson ratio (R) from the relationship $R = \frac{\pi^2 k_B^2}{3\mu_B^2} \cdot \frac{\chi_0}{\gamma}$. The Wilson ratio, a dimensionless parameter concerned with correlations among electrons, is equal to or about 1 for metals and normal compounds, and increases as the *e-e* interaction increases. We obtain R = 1.80, which implies an enhanced *e-e* interaction in SmOFeAs.

Other measurements in literature [18] reported a higher $\gamma$ value of 119.4 mJ/K$^2$ mol. With this value an unphysical low value of R = 0.63 is obtained, which casts some doubt on these data. In the undoped La-1111 compound a estimate of $\chi_0$ gives a maximum value of 5$^{.}$10$^{-5}$ emu/mol (see Fig.4). Using a value of 3.7 mJ/ K$^2$ mol for $\gamma$ [9]  we obtained the maximum value of 1 for the Wilson ratio, as expected for systems with free electron character.

In the left side inset of Fig.4 the low temperature part of $\chi$ versus T data, measured for different values of the magnetic field, is shown. It is noticeable that no variation of the Sm ions antiferromagnetic ordering temperature is observed up to the maximum field of 50 kOe. The independence of also very low antiferromagnetic ordering temperatures under strong applied magnetic fields is rare, but it has been already observed in rare earths and actinides compounds. For example, CeRhIn$_5$ displays the AF transition at 3.8 K and the same transition is still detected at the same temperature with a magnetic applied field of 90 kOe [23]. Another example is PuPd$_2$Sn [24], where both magnetisation and heat capacity measurements detected the same AF transition temperature ( $T_N$ = 11 K) at 0 and 90 kOe applied magnetic fields. In the same inset the low temperature resistivity is reported: the agreement between the susceptibility data, displaying the AF peak at 6K, and the resistivity values, where the strong slope change begins at the same temperature, is remarkable.

Finally we underline that the overall magnetic behaviors are in agreement with the resistivity curves presented in Fig.2. In both 1111–compounds the high T magnetic orderings (SDW) correspond to resistivity decreases at comparable temperatures.

### 3.2 Doped samples

### 3.2.1 Normal state

In Fig.5 molar susceptibility versus temperature measurements are presented for doped La and Sm-1111 compounds from T=2K up to T= 360K. The applied magnetic field was 30kOe for ZFC and FC procedure.

The main features in Fig.5, common to both the samples, are the following: i) a clear decrease in magnetization, indicating the superconducting transition, can be seen at $T_c$~19K and $T_c$~50K for La and Sm-1111 samples, respectively, ii) above $T_c$ a monotonous decrease in magnetization with increasing temperature (Curie-Weiss type) is observed up to about 330K, after which a slight increase in magnetization with a faint maximum at T~350K occurs (the up right side inset shows an enlargement of the molar susceptibilities in the high temperature region where the maximum at about T=350K is evidenced), and iii) no feature related to the SDW is observed. For Sm-1111 compound only, a complex behaviour in ZFC-FC measurements on the superconducting side may be observed, which we will discuss in the next paragraph.

The field dependence of the magnetization measured at T=30 K (just above the superconducting transition) and T=300K for La-1111 sample and at T=60 K and T=300K for Sm-1111 sample are shown in the bottom left side and bottom right side insets, respectively. Different behaviors are observed. For SmFeAsO$_{0.85}$F$_{0.15}$ at both temperatures the magnetic signal increases linearly with the field, which indicates that no ferromagnetic background is present, but the linear slope is different and higher than that observed in the undoped sample. In the LaFeAsO$_{0.85}$F$_{0.15}$, on the contrary, a concavity due to the presence of ferromagnetic impurities is observed both at 30K and 300K. In particular, at T=30K the continuous variation of slope suggests the presence of ferromagnetic clusters. Also in this sample the linear part of the signal is highly increased respect to the undoped sample.

In doped La-1111 compound it has been shown that the intrinsic susceptibility of Fe ions exhibits a linear increase with temperature [13,15], as we observed in the undoped compound above the SDW transition. Therefore we hypothesize that the Curie-Weiss type behaviour, observed in the normal state of the $LaFeAsO_{0.85}F_{0.15}$ sample, could be due to the presence of paramagnetic impurities.

In $SmFeAsO_{0.85}F_{0.15}$ sample a Curie-Weiss type behavior is expected, due to the presence of magnetic Sm ions: but, as previously outlined, the Curie-Weiss type signal increased by about a factor two compared to the undoped one (see Fig.4 and Fig.5 for the comparison), which indicates that also in this sample a great part of the magnetic signal is spurious and paramagnetic.

In an effort to study in more detail the spurius phases present in our doped samples we prepared and measured SmOF and FeAs, while magnetic contributions eventually arising from $Fe_2As$ and $FeF_2$ were estimated from literature data ($FeF_2$ is a magnetic compound used as precursor in the sample preparation). SmOF exhibits a paramagnetic susceptibility with a room temperature value of about $10^{-3}$ emu/mol and FeAs exhibits a signal of the order of $10^{-3}$ emu/mol nearly constant with temperature. $Fe_2As$ is known to order antiferromagnetically at T=353 K [16]; on the contrary, the detailed magnetic characterization performed on $FeF_2$ [25,26] reveals that this compound is antiferromagnetic below $T_N = 78$ K and the Curie-Weiss behaviour presented above is ascribed to an iron magnetic moment of 4.7 $\mu_B$.

On the basis of the afore mentioned quantitative Rietveld analysis we subtracted a contribution of 4.6% SmOF and 5.6% FeAs from the experimental data. In addition, since the peak in both samples at T=350K, (inset of Fig.5) suggested to us the presence of $Fe_2As$, we tried to calculate its contribution. However, the literature data on this phase [16] are affected by the presence of ferromagnetic impurities, thus preventing any quantitative estimate. Finally, we hypothesized that the strong paramagnetic signal present in both $LaFeAsO_{0.85}F_{0.15}$ and $SmFeAsO_{0.85}F_{0.15}$ samples could be ascribed to $FeF_2$. However, the estimate of $FeF_2$ concentration to simulate the experimental susceptibility curve correctly, leads to an unrealistic content higher than 6%, but this phase has not been detected neither by XRPD nor by SEM analysis. It seems that probably the contribution of some other spurious phases, in concentration lower than XRPD or SEM threshold detectability level, must be taken into account. In conclusion the normal state susceptibility of the doped-1111 phases could not be extracted.

### 3.2.2 Superconducting state

As it may be seen in Fig.5 the behaviour of LaFeAsO$_{0.85}$F$_{0.15}$ sample below the superconducting transition temperature T$_c$ is the standard diamagnetic one (with ZFC and FC curves nearly coincident which indicates a large reversibility region in the H-T phase diagram), while the behavior of the SmFeAsO$_{0.85}$F$_{0.15}$ sample is quite unusual. In Fig.6 an enlargement of Fig. 5 in the low temperature region is shown for the data relative to the doped Sm-1111 sample, to observe its superconducting behavior in detail: the four curves correspond to ZFC-FC measurements made starting from T=2K (open symbols) or from T=5K (filled symbols). At T~50K magnetization begins to decrease, owing to the diamagnetic signal of the superconducting state, but, as temperature is lowered, a minimum occurs after which the signal increases. The ZFC and FC measurements are coincident from T$_c$ down to about the temperature of the minimum (~25 K), which indicates also for this sample a large zone of reversibility in the H-T phase diagram. At lower temperatures ZFC and FC curves separate, but measurements performed starting from 2K or 5K do not overlap, showing different behaviors that depend on the measurement starting temperature. In the FC curves a slightly higher value is observed for the measurement that started from 5K. Moreover, the curve starting from 2K shows a faint maximum at 4K. The ZFC curves show the expected shielding, but with a higher susceptibility values for the measurement that started from 2K. These differences are small but outside the experimental error. Such phenomenology may be understood bearing in mind that Sm ion sublattice orders antiferromagnetically at low temperature, as we have observed in the parent sample (see Fig.3). Such ordering has been evidenced by Tropeano et co-workers [23] in both SmFeAsOF and SmFeAsO$_{0.85}$F$_{0.15}$ samples by specific heat measurements. In particular, in SmFeAsO sample a peak at T= 4.6 K is observed, while in the superconducting one the peak is shifted to T= 3.7 K. Below this temperature superconductivity and RE lattice antiferromagnetism coexist. The ZFC and FC curves originated at T=2K show less irreversibility (i.e. are nearer) because, starting from a temperature just below T$_N$ and passing at T$_N$ through the maximum of magnetization of the Sm sublattice, the superconductor feels a higher internal magnetic field: that is the magnetic ordering affects the superconducting properties. The slightly increasing FC signal observed from 2K up to 5K depends on the competition between the superconducting FC signal and the magnetic signal of Sm ions sublattice. To study in more detail the evolution of the competition-coexistence between superconductivity and magnetism we performed FC measurements at magnetic field higher than 30 kOe .

The results are presented in Fig.7. For sake of clarity we show the magnetization, instead of susceptibility: thus, the measurements at different fields are spaced and better visible. The increase of the magnetic field produces a less pronounced diamagnetism and makes the antiferromagnetic peak of Sm ions more and more visible. The position of the peak agrees with the maximum observed in the specific heat measurements.

In conclusion, Sm antiferromagnetism and superconductivity appear to coexist in $SmFeAsO_{0.85}F_{0.15}$ sample at low temperature. Since antiferromagnetic ordering of the RE ions bearing magnetic moments in iron pnictides has been observed for other RE (Nd,Ce,Pr,Gd), this coexistence is a general character of this new class of superconducting compounds, as of many other classes of superconducting and magnetic materials ($RERh_4B_4$, borocarbides, ruthenocuprates, ….). We recall that superconductivity and RE magnetism are located in different planes of the unit cell in $SmFeAs(O_{1-x}F_x)$. On the other hand, the problem of the magnetism in the FeAs planes, where superconductivity sets in, is a totally open question

## 4.Conclusions

We have prepared and characterized undoped and 15% F doped La and Sm-1111 compounds.

In the undoped samples we estimated the temperature independent Pauli paramagnetic susceptibility and therein the Wilson ratio: while in La-1111 phase the obtained value indicates a nearly free electron character, in Sm-1111 compound a heavy fermions character is observed. Moreover, in this last sample, by subtracting the Sm paramagnetism from the total signal, we estimated the amplitude of the susceptibility step at the SDW temperature, that turned out to be comparable with that of the La one. In the doped samples the superconducting behavior could be observed, and for Sm-1111 phase the competition-coexistence of superconductivity and antiferromagnetic ordering of Sm ions sublattice was clearly detected.

## Acknowledgment

Financial support of the "Compagnia di S. Paolo" and of the Italian Foreign Affair Ministry (MAE)-General Direction for the Cultural Promotion and Cooperation is acknowledged

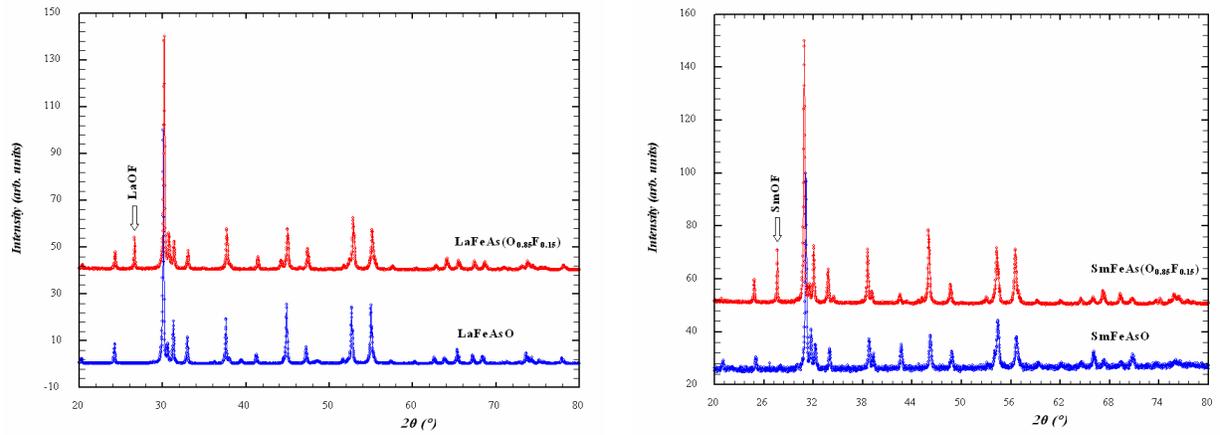

Fig.1

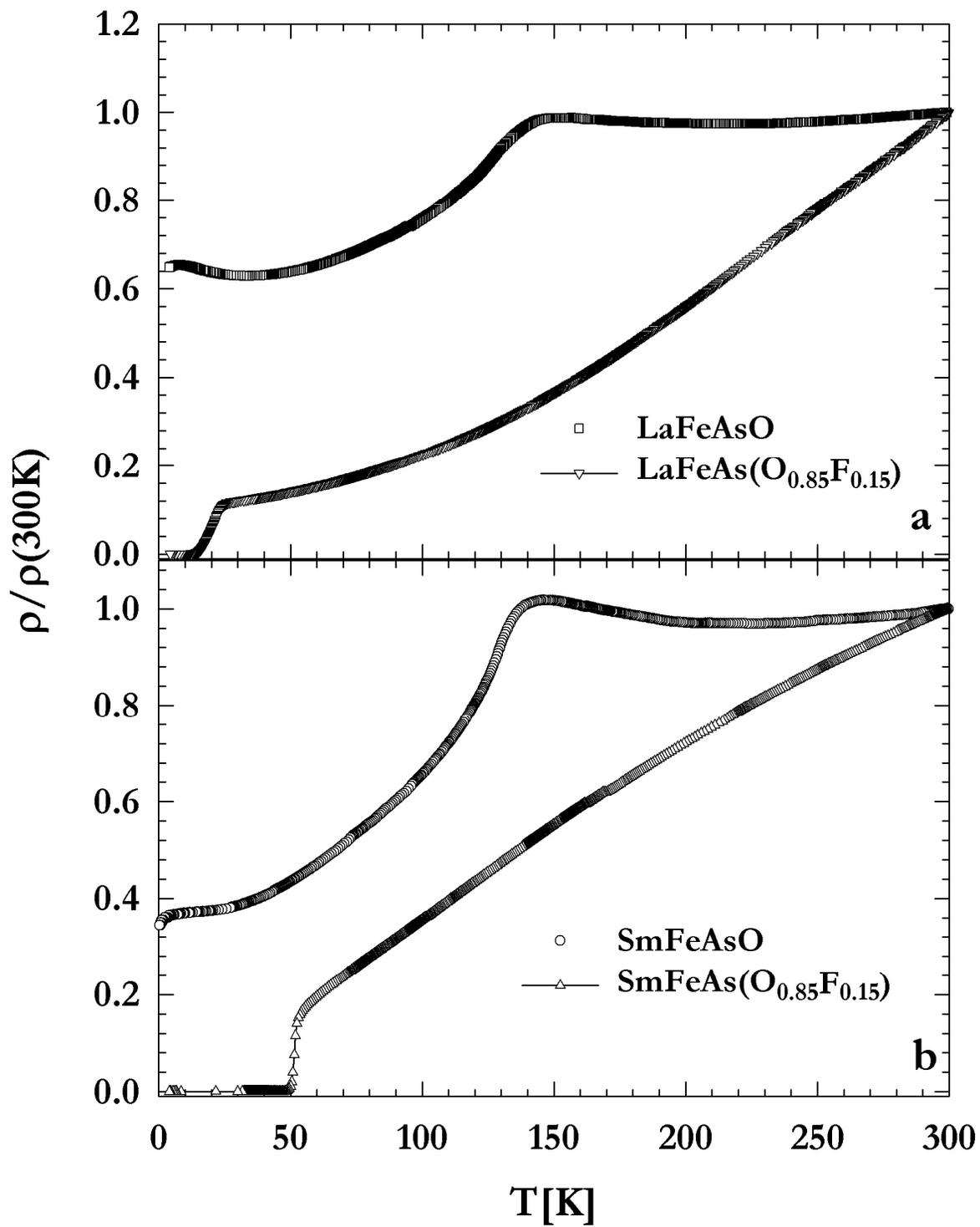

Fig.2

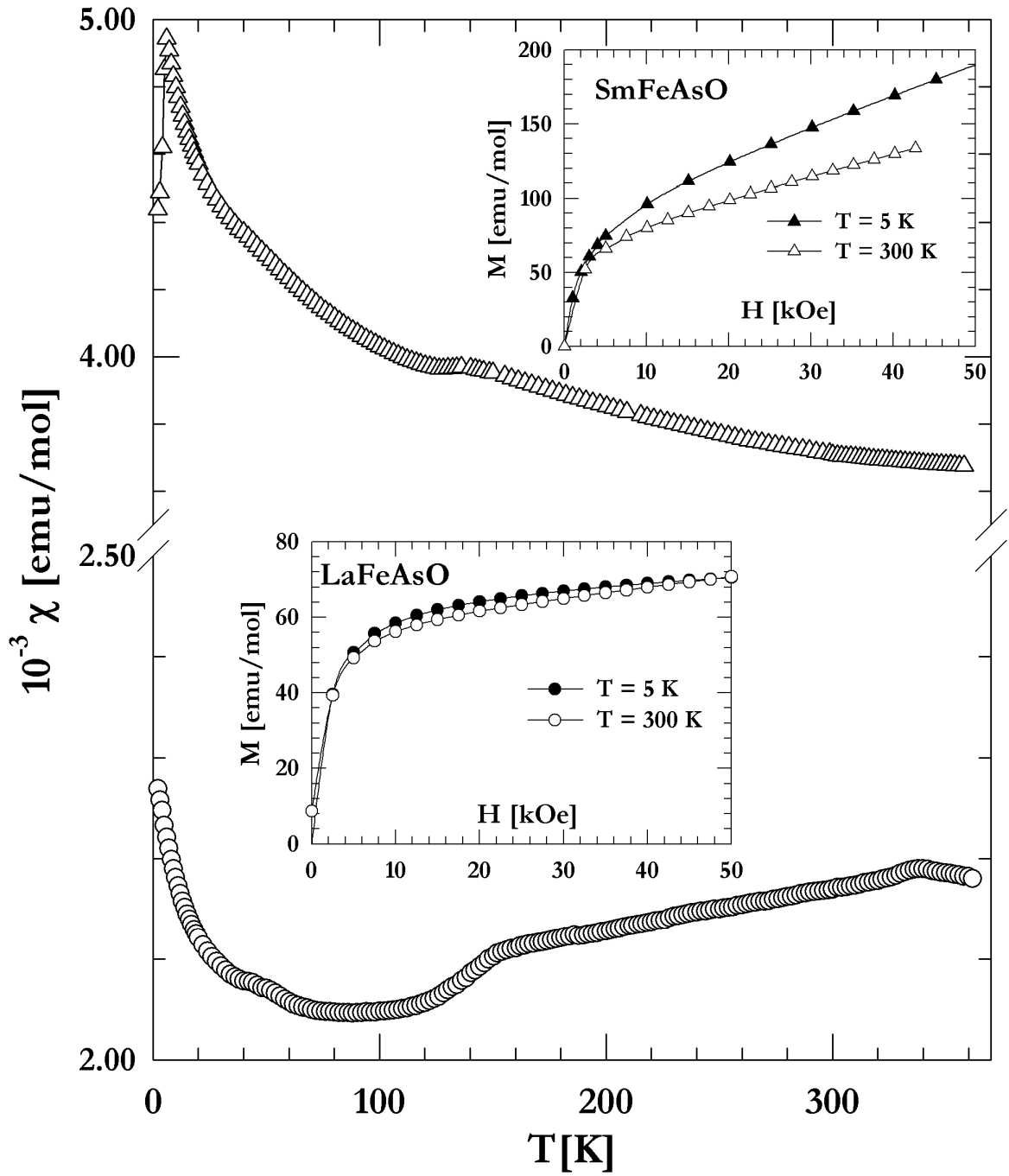

Fig.3

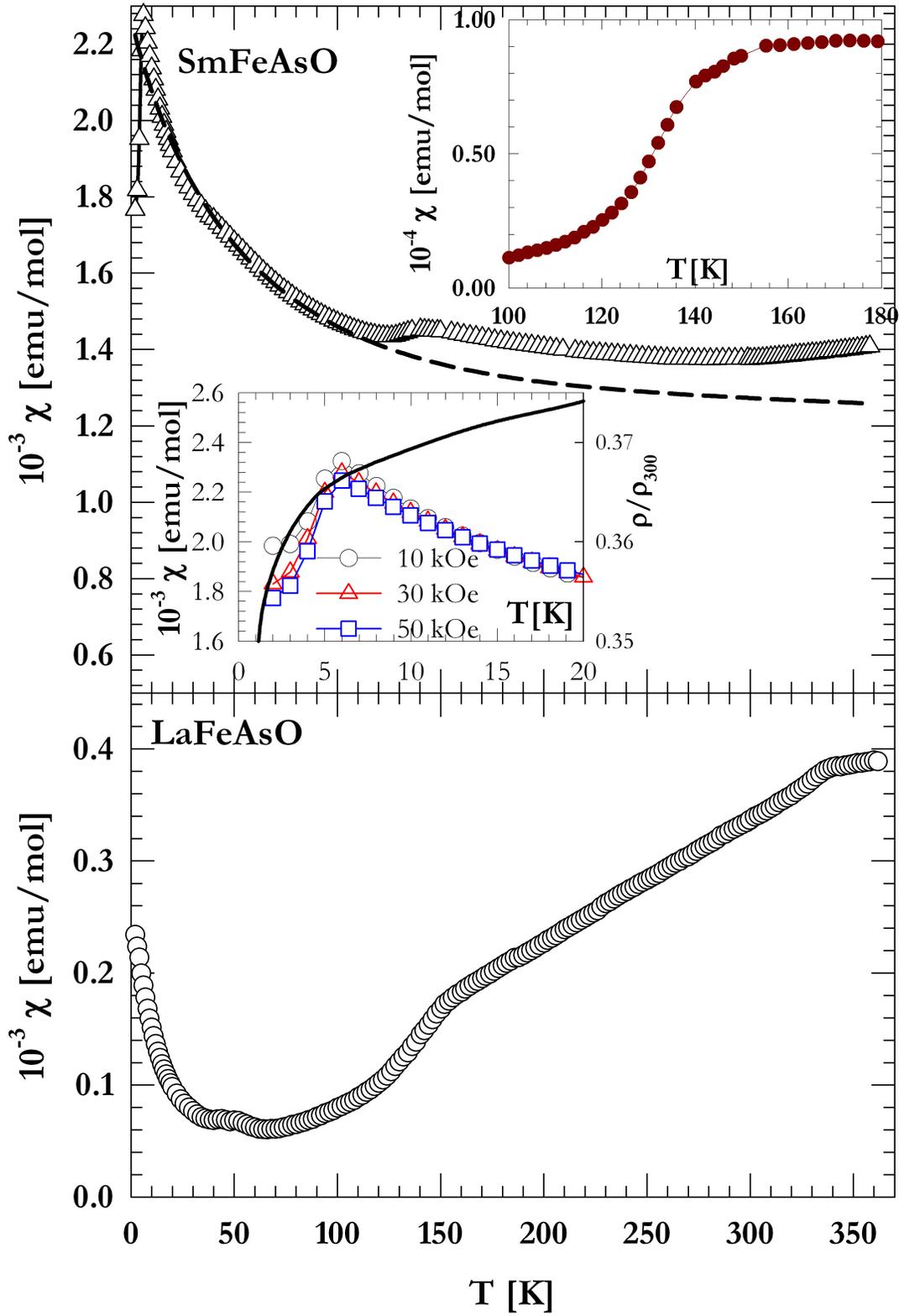

Fig.4

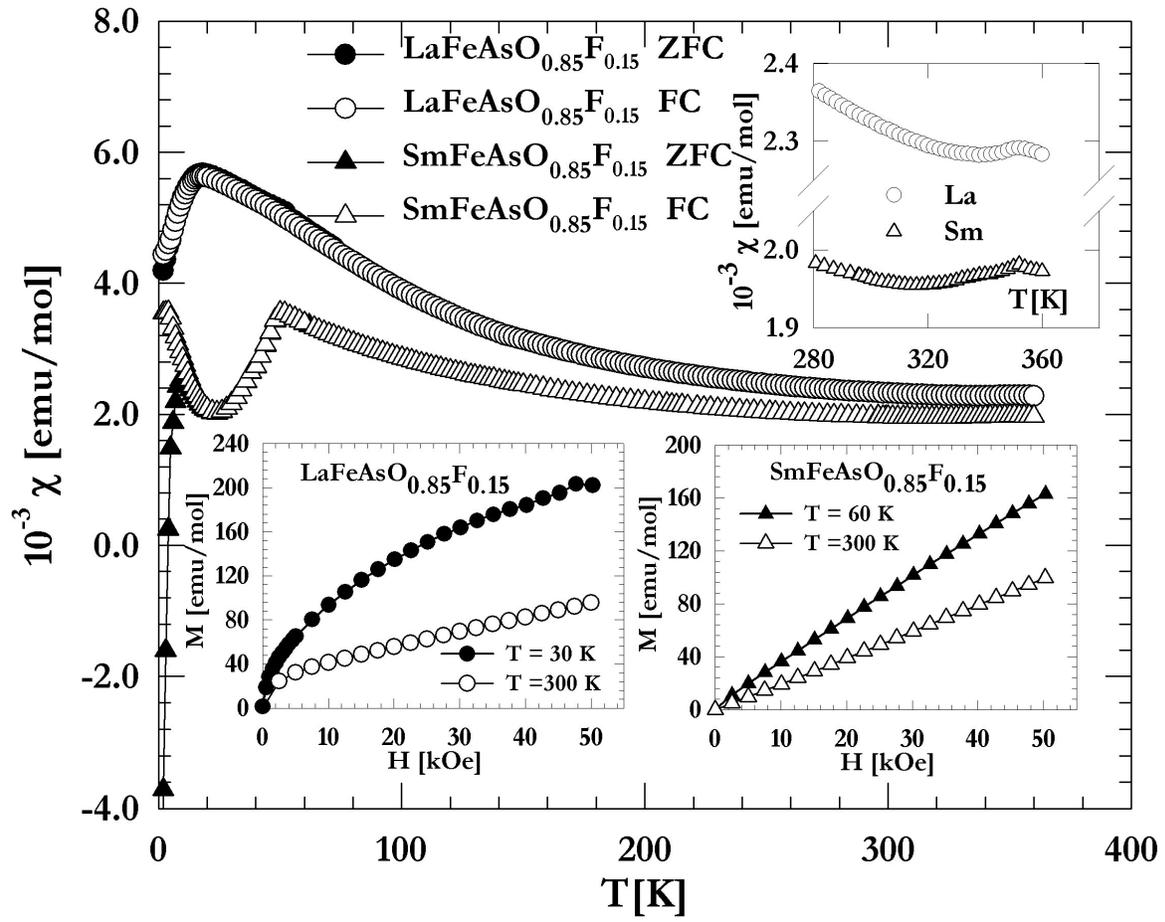

Fig.5

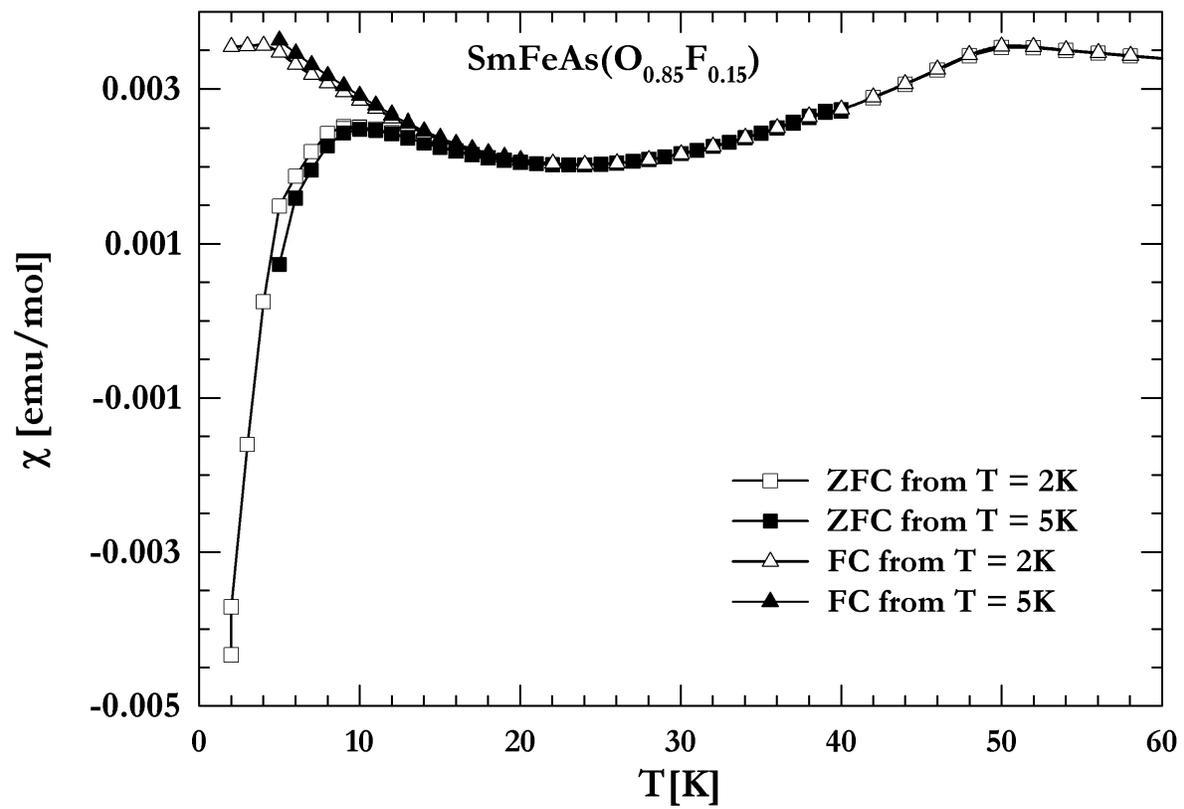

Fig.6

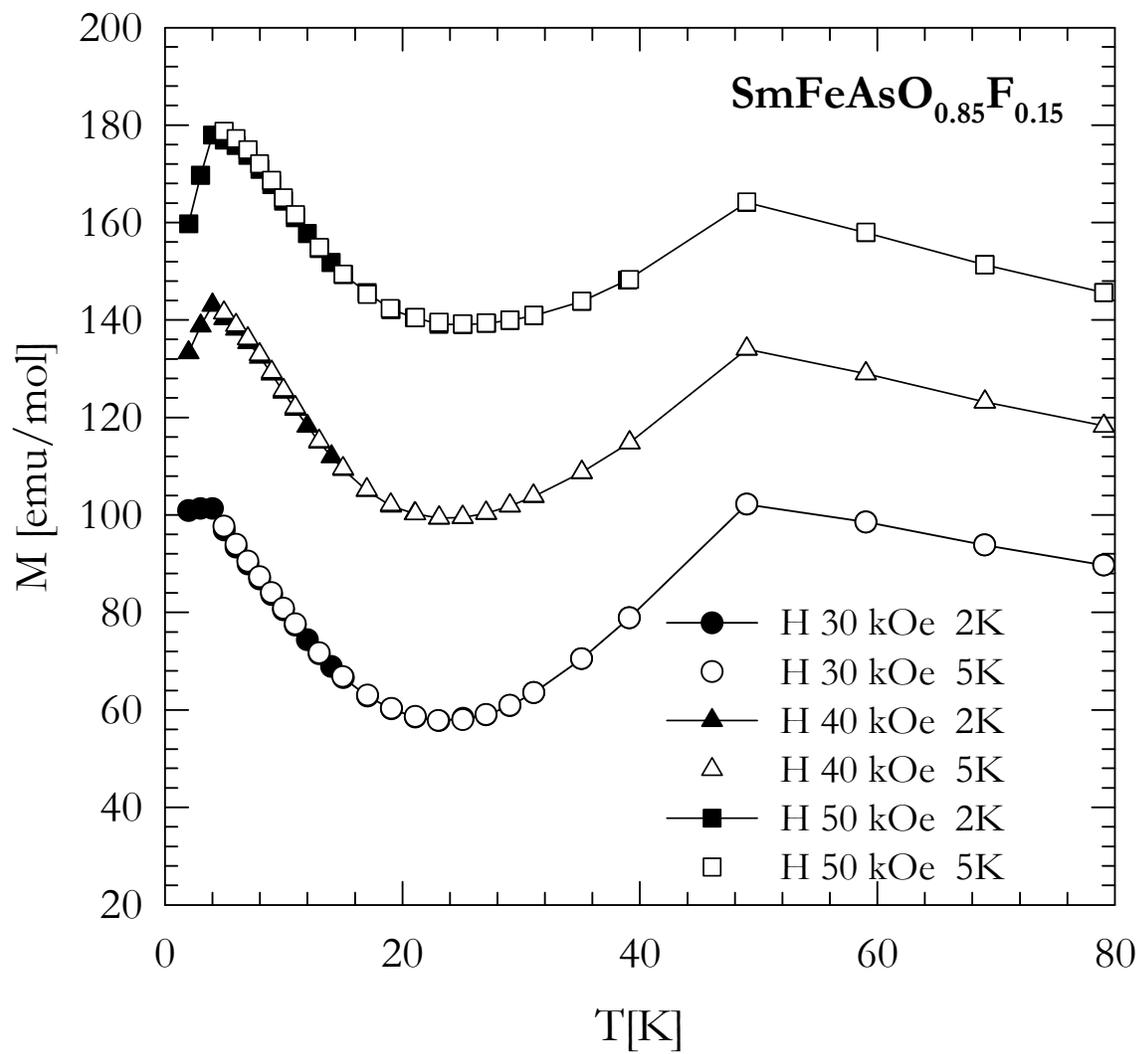

Fig.7

**Figure caption**

**Fig.1** Comparison between the XRPD patterns of the pure and 15% F doped LaFeAsO (on the left) and SmFeAsO (on the right) oxy-pnictides; both F-doped samples are characterized by the presence of the respective rare earth oxy-fluoride (arrowed).

**Fig.2** Resistivity versus temperature measurements of undoped and 15%F doped LaFeAsO and SmFeAsO compounds.

**Fig.3** Molar susceptibility versus temperature of LaFeAsO and SmFeAsO samples in the range 2K-360K. The applied field is 30kOe. Upper right inset: magnetization versus field at T=5K and T=300K for SmFeAsO. Lower left inset: magnetization versus field at T=5K and T=300K for LaFeAsO

**Fig.4** Molar susceptibility versus temperature corrected for the ferromagnetic background (see text) of LaFeAsO and SmFeAsO samples in the range 2K-360K. The dotted line is the fit of Sm ions susceptibility. Upper right inset: intrinsic susceptibility contribution of the Fe-As layers. In the lower left inset $\chi$ versus T data for SmFeAsO at low temperature and different values of the applied magnetic field are presented (left Y-scale) together with the low T resistivity data (continuous line) on the right Y-scale.

**Fig.5** Molar susceptibility versus temperature of $LaFeAsO_{0.85}F_{0.15}$ and $SmFeAsO_{0.85}F_{0.15}$ compounds. The applied field is 30kOe. Upper right inset: molar susceptibilities in the high temperature region. Lower right inset: magnetization versus field at T=60K and T=300K for $SmFeAsO_{0.85}F_{0.15}$. Lower left inset: magnetization versus field at T=60K and T=300K for $LaFeAsO_{0.85}F_{0.15}$.

**Fig.6** Molar susceptibility versus temperature of $SmFeAsO_{0.85}F_{0.15}$ in the low temperature region. Open symbols: ZFC-FC measurements performed starting from 2K; filled symbols: ZFC-FC measurements performed starting from 5K. The applied field was 30kOe.

**Fig.7** FC magnetization versus temperature of SmFeAsO$_{0.85}$F$_{0.15}$ in the low temperature region for different applied fields starting from 2K (full symbols) and 5K (open symbols).